\documentstyle[pra,aps,epsfig,multicol]{revtex}

\newcommand{\beq}{\begin{eqnarray}}
\newcommand{\eeq}{\end{eqnarray}}
\newcommand{\la}{\langle}
\newcommand{\ra}{\rangle}
\newcommand{\intot}{\int_{0}^{T}}
\newcommand{\sit}{\psi(t)}
\newcommand{\siT}{\psi(T)}
\newcommand{\sid}{\dot{\psi}{t}}
\newcommand{\pit}{\lambda(t)}
\newcommand{\piT}{\lambda(T)}
\newcommand{\pid}{\dot{\lambda(t)}}

\begin{document}
\draft

\title{Optimally Shaped Terahertz Pulses for Phase Retrieval in a Rydberg Atom Data Register}

\author{C.~Rangan and P.~H.~Bucksbaum}
\address{Physics Department, University of Michigan, Ann Arbor, MI 48109-1120}

\maketitle

\begin{abstract}
We employ Optimal Control Theory to discover an
efficient information retrieval algorithm that can be performed
on a Rydberg atom data register using a shaped terahertz pulse.
The register is a Rydberg wave packet  with one consituent orbital
phase-reversed from the others (the ``marked bit'').  The
terahertz pulse that performs the decoding algorithm does so by
by driving electron probability density into the marked orbital.
Its shape is calculated by modifying the target of an optimal
control problem so that it represents the direct product of all
correct solutions to the algorithm.
\end{abstract}


%\begin{multicols}{2}
%\narrowtext
\section{INTRODUCTION}

The interaction of terahertz frequency electromagnetic pulses with Rydberg
atoms has produced many insights into the dynamical properties of atomic
systems~\cite{thzpapers}.  The comparable time scales of terahertz pulses with
those of Rydberg state lifetimes make it possible to envision schemes of
quantum control.  In this paper, we propose a method of controlling Rydberg
wave packets using shaped terahertz pulses, and theoretically show how these
pulses can be designed to execute a quantum algorithm on a Rydberg atom
data register.

It has been shown that information can be stored in the phases of the
constituent orbitals of a Rydberg wave packet~\cite{ahnSc00,redist}.  Recently,
a terahertz half-cycle pulse was used to decode the information stored in
a Rydberg atom data register~\cite{redist}.
This half-cycle pulse decodes the phase structure by
retaining the population only in the orbital that was initially
$180^{\circ}$ out of
phase with respect to the other orbitals, i.~e.~, the marked bit.
However, this guess (unshaped) pulse does not decode all marked bits
of the register with the same efficiency.
In this paper, we aim to
find the shaped THz pulse that will optimally transfer most of the population
to any marked bit of the quantum data register.

\section{OPTIMAL CONTROL THEORY}
To design the terahertz pulse, we use a method that has been used
extensively in mathematical and engineering applications ---
Optimal Control Theory (OCT)~\cite{krotovbook,blochbook}.
This theory has also been applied to the control of quantum
systems~\cite{piercePRA88,shiJCP90,kosloffCP89,jacubetzCPL90,somloiCP93,%
kimPRA95,yanCP97} with some success in experimental implementation~\cite{judsonPRL92}.  We
use OCT to design a terahertz frequency pulse that can be
used to achieve a desired target state. We then modify the OCT
target state to make it possible to discover not a single target,
but an optimized quantum algorithm. The wave function of the
Rydberg electron is the state variable, and the electric field of
the terahertz pulse is the control parameter.  A functional $J$ is
defined, whose extremum must be calculated. The functional
consists of two parts, representing the desired target and the cost. 
Our aim is to maximize the
fraction of the electron probability density in a target orbital
$|a_k \ra $ at a time T (after the end of the terahertz pulse).
That is, the target functional $ \la P_k (T)\ra=\la \psi(T)|a_k\ra \la a_k | \psi(T)
\ra$ must be a maximum. The cost functional represents the
constraint on the control parameter,
 the terahertz field $E(t)$.
The integrated energy of the pulse must be kept low, therefore the cost functional is defined as
$Y(T)=\intot dt \ \ell(t)|E(t)|^2$.
Here $\ell$ is a penalty parameter, in general
time dependent, that controls the cost functional, and hence the peak
terahertz field.
The functional $J$ written as
\beq
J & = & \la \psi(T)|P_k|\psi(T) \ra  -  Y(T),
\eeq
\noindent must be maximized.

The aim is to find an optimal control function $E(t)$ that maximizes $J$.
The wave packet evolution is
governed by the Schr\"{o}dinger equation.
In atomic units,
 $e=m_e=\hbar = 1$,
\beq
|\dot{\psi}(t)\ra & = & -i H(t)|\psi(t) \ra , \label{eq:sch}
\eeq
where $H(t)=H_0+E(t)z$.   The equation of motion acts as a constraint
on the evolution of the state, and in a manner similar
to that used in variational calculus,
we introduce a Lagrange multiplier $|\lambda(t)\ra$~\cite{gerjuoyRMP83}.  The
unconstrained functional that must be optimized is written as
\beq
\bar{J} & = & J - \intot dt \left[ \la\pit|\sid\ra+\la \pit |iH|\sit \ra +
\la \sit|\pit \ra \right. \nonumber \\
&   & \left. -i \la \sid|H|\pit \ra \right]. \\
& = & \la \siT | P_k | \siT \ra
 - \intot dt \ \ell(t) |E(t)|^2 
- 2 Re \la \pit | \sit \ra |_0^T \nonumber \\
&   & 
+ \intot dt \ 2 Re \la \pid | \sit \ra
- \intot dt \ 2 Re \la \pit | iH | \sit \ra.
\eeq

Several iterative techniques for determining the optimal solution have been 
developed~\cite{shiCPC91,krotovEC83,kazakovARC87,tannorbook,zhuJCP98}.  
Following the scheme for an iterative solution proposed in
Ref.~\cite{tannorbook}, the functional $\bar{J}$ is written as the sum of a
terminal part and an integral
\beq
\bar{J} & = & G+\intot dt \ R, {\rm where}\\
G & = & \la \siT|P_k | \siT \ra - 2 Re\la \pit|\sit \ra |_0^T, \\
R& = & - \ell(t) |E(t)|^2 + 2 Re \left[ \la \pid | \sit \ra - i\la \pit |H| \sit \ra \right] .
\eeq
\noindent The maximum of both G and R is sufficient to ensure the maximum of $\bar{J}$.
Our objective is to iteratively determine the optimal function $E(t)$ that maximizes $\bar{J}$.
The functional $\bar{J}$ at the $k^{th}$ and $(k+1)^{st}$ iteration is
defined by the changes in the wave function and the field.  That is,
\beq
|\sit^{k+1} \ra & \equiv & |\sit^k \ra + |\Delta \sit \ra, \nonumber \\
E^{k+1} & \equiv & E^{k}+\Delta E.
\eeq
The difference in $\bar{J}$ between two successive iterations is written as
\beq
\bar{J}^{k+1} - \bar{J}^{k} & = & \Delta_1 + \Delta_2 + \Delta_3, \ {\rm where} 
\eeq
\beq
\Delta_1 & \equiv & G(\psi^{k+1}(T)) - G(\psi^k(T)) \nonumber \\
& = & 2 Re \left[ \la \psi^k(T)|P_k|\Delta \psi (T) \ra 
- \la \piT|\Delta \siT \ra \right] \nonumber \\
&   & + \la \Delta \siT |P_k |\Delta \siT \ra, \\
\Delta_2 & \equiv & \intot dt [R(t, \psi^{k+1}, E^{k+1}) - R(t, \psi^{k+1}, E^k)] \nonumber \\
& = & -2 Re \intot dt \ \ell(t) E^{\ast k}(t) \Delta E(t) \nonumber \\
&   & - \intot dt \ \ell(t) |\Delta E(t)|^2 \nonumber \\
&   & -2 Re \intot dt \left[ i \la \lambda^k(t)|z \Delta E(t)| \psi^k(t) \ra \right. \nonumber \\
&   & \left. + i \la \lambda^k(t)|z \Delta E(t)|\Delta \sit \ra \right] , \\
\Delta_3 & \equiv & \intot dt [R(t, \psi^{k+1}, E^k) - R(t, \psi^k, E^k)] \nonumber \\
& = & 2 Re \left[ \intot dt \la \pid | \Delta \sit \ra \right. \nonumber \\
&   &  \left. - i\intot dt  \la \pit | H^k| \Delta \sit \ra \right] .
\eeq

\noindent
Choosing
\beq
| \piT \ra & = & P_k |\psi^k(T) \ra, {\rm and} \label{eq:c1}\\
| \pid \ra & = & -i H^k |\pit \ra, {\rm we \ find} \label{eq:c2}  
\eeq
\beq
\Delta_1 & = & \la \Delta \siT | P_k | \Delta \siT \ra . \\
\Delta_2 & = & -2 Re \intot dt \left[ \ell(t) E^{\ast k}(t) \Delta E(t) + \frac{1}{2} \ell(t) |\Delta E(t)|^2 \right. \nonumber \\
&   & + \left. i \la \lambda^k(t)|z \Delta E(t)| \psi^k(t) \ra + \right. \nonumber \\
&   & \left. i \la \lambda^k(t)|z \Delta E(t)|\Delta \sit \ra \right] . \\
\Delta_3 & = & 0 .
\eeq
\noindent The solution
improves when $\bar{J}$ increases or stays the same at each iteration.  $P_k$
is a positive semidefinite operator, therefore $\Delta_1$ is greater than or equal to
zero.  The change in the control parameter $\Delta E$ is chosen to
maximize $\Delta_2$.  
The expression for $\Delta_2$ suggests the following 
change in the field at the
(k+1)th iteration~\cite{tannorbook}: 
\beq
\Delta E(t) & = & \frac{-i}{\ell(t)} \la \lambda^{k}(t)|z|\psi^{k+1}(t) \ra. \label{eq:c3}
\eeq
The appearance of $|\psi^{k+1}(t) \ra$ in the above expression implies that the
overlap of $|\sit \ra$ and $| \pit \ra$ is fed back immediately to find the field at
the next time step.

The optimal control algorithm consists of the following steps:
\begin{enumerate}
\item
Starting from the initial wave packet $|\psi^{(0)}(0) \ra  = |\psi(0) \ra $ and a first guess for
the terahertz field $E^{(0)}(t)$, the wave packet is
propagated according to Eq.~\ref{eq:sch} to find $| \psi^{(0)}(T) \ra$.

\item
Using Eq.~\ref{eq:c1} to find $|\piT \ra$, Eq.~\ref{eq:c2} is iterated backward
to time $t=0$, and $|\pit \ra $ is found at every time step.

\item
With $|\psi^{1}(0) \ra = |\psi(0) \ra$, equation~\ref{eq:c3} is then used to
find a new value of the control field $E^{1}(t)$ and equation~\ref{eq:sch} is used to propagate the wavepacket forward in time.
\end{enumerate}
The second two steps are repeated until the target yield converges to within the
desired accuracy.

\section{OPTIMAL PULSE FOR A SINGLE TARGET STATE}
The first step in our approach is to find the optimal THz field
required to decode a single flipped state.  The initial state of
the Rydberg data register is a wavepacket made of the $24p$ through $29p$
orbitals of equal amplitudes and the phase of the $26p$ (the
marked bit) orbital opposite to that of the others.  
The initial
guess THz pulse is a half-cycle pulse~\cite{shakeshaft} of pulse
width $1ps$.  The desired initial phase structure occurs at the
peak of the half-cycle pulse (i.~e.~at $0.5$ps).  
We find the THz pulse that will
optimally transfer most of the population to the marked bit. The
best value of the penalty parameter, $\ell$, that controls the
peak field of the THz pulse to a reasonable value, and at the
same time produces the desired final state, was found to be
roughly $10^{10}$. One feature of several optimal fields obtained 
theoretically~\cite{shiJCP90,kimPRA95} is that the fields do not go to zero 
smoothly at the times $t=0$ and $t=T$.  
In the present system, the correct evolution of the wave 
packet depends very sensitively on the fields at the end points.
The addition of a smooth switch-on and switch-off of the calculated optimal 
field drastically changes the evolution of the wave packet.  Therefore, 
the condition that the THz field continuously goes to zero before and after 
the time interval of choice must be built into the algorithm.
To ensure that the THz field goes smoothly to
zero at times $t=0$ and $t=T$, the penalty parameter $\ell(t)$ is made
a smoothly varying time-dependent function.  The penalty on the 
pulse fluence is a thousand times more at the end points than at the rest
of the pulse duration.  The smoothness of the penalty function ensures a 
smooth switching on and switching off of the THz pulse.

This OCT implementation is very successful in describing the
terahertz control of a Rydberg wave packet.  The method takes
macrosteps in the control field at every iteration and
convergence is swift. The computational complexity is of the same
order as that of the wave packet propagation.  Therefore, we use
a split-operator method in a restricted basis of essential states
\cite{feit,redist}.  
The energy eigenstates of 
cesium are calculated using a pseudopotential method on a nonlinear radial 
grid~\cite{grid}.
A time step of 10fs ensures the accuracy of
the propagator, which is correct through the second-order in the
time step. The numerical implementation of the local iterative
algorithm is extremely sensitive to numerical error, and
$\Delta_3$ must be maintained equal to zero to very high
precision \cite{tannorbook}. The unitary nature of the
symmetrized-product propagator maintains this condition. The
restricted basis consists of 195 energy eigenstates with
principal quantum number $n$ ranging between 21 and 31, and
angular momentum quantum number $\ell<17$. Absorbers ensure that
population does not get reflected from the $n=21$, $n=31$, and
$\ell=16$ `boundaries'.  Using the selected state basis also
imposes the condition that the spectrum of the THz pulse is
decided by the energy range of the selected state basis.  In this
implementation of optimal control theory, we have chosen a fixed
pulse length, $T$, of roughly 8ps.  In other formalisms, this
time $T$ may also be varied as a parameter.

The THz field that optimizes the population in the marked $26p$ state is shown
in Fig.~\ref{fig1}(a).  The initial population in the $26p$ state is $16.7\%$.  The optimal
pulse will decode the information stored as phase by transferring most of the
population into the $26p$ state.  With the
initial guess pulse, the population is $29.5\%$.  After 50 iterations, the
target yield is increased to $52.8\%$.  The spectrum and Husimi
distribution~\cite{husimi}
of this optimal pulse are shown in Fig.~\ref{fig1}(b) and
Fig.~\ref{fig1}(c) respectively.  Notably, the strong peaks in the spectrum and the
Husimi distribution do not correspond to any resonance between the energy
levels of the selected state basis.  The optimal terahertz pulse does not
drive the system to any particular resonant condition.  Instead, it alters
the phases of the constituent orbitals of the wave packet so that they
interfere
to produce the desired probability distribution.

Figure~\ref{fig2} shows the evolution
of the wavepacket as a function of time while the optimal pulse is on.
During the THz pulse,
probability can leak into other states not in the register (the other
states in the essential basis).
At the end of the pulse, a large fraction of the electron probability density
lies in the flipped orbital (marked bit) of the data register.
This can be thought of as using the other states of the data register as
working qubits, which are used during the computation, but are not
measured for any useful retrieval of information.

One interesting feature of this optimal pulse is that the peak field of
roughly $1{\rm KV/cm}$ lasts for roughly $0.5$ps.
For a $\bar{n}=26$ wavepacket,
this field which is beyond the field ionization limit lasts for more than half
the Kepler period ($\sim 2 \pi n^3$).  Yet, $99\%$ of the population remains in
the selected state basis.  This feature is an example of interferometric
stabilization~\cite{fedorov}, seen in other atomic systems.
 
\section{OPTIMAL PULSE FOR A QUANTUM ALGORITHM}

This THz pulse is optimal only for decoding the flipped $26p$
orbital. That is, if the phase of a different state were flipped,
this pulse will not decode it. We wish to design a universal THz
pulse that will optimally decode any flipped orbital of the wave
packet register.  Therefore we redefine the optimal control
problem by considering an initial state that is a product state
of independent wave packets with singly flipped orbitals. 
\beq
|\Psi(0) \ra & = & |\psi^{(1)}_{25p}(0)\ra \otimes
\psi^{(2)}_{26p}(0) \ra \otimes \psi^{(3)}_{27p}(0) \ra \otimes
\cdots 
\eeq 
The terahertz pulse acts simultaneously, but
independently on all these wave packets.  The desired final state
is also a product state of independent wave packets with the
flipped bit correctly decoded. 
\beq |\Psi(T)\ra & = & |25p^{(1)}
\ra \otimes |26p^{(2)} \ra \otimes |27p^{(3)} \ra \otimes \cdots
\eeq 
The counterparts of Eq.~\ref{eq:c1} and Eq.~\ref{eq:c2} are
straightforward. At every time step, the updated THz field is
found by using a modified version of Eq.~\ref{eq:c3}, with the
matrix element of $z$ replaced by a sum of matrix elements of
$z$, one from each independent `subspace'. 
\beq \Delta E(t) & = &
\frac{-1}{\ell(t)} \sum_{i=1}^N \la \lambda^{k}_{(i)}
(t)|iz|\psi^{k+1}_{(i)}(t) \ra. 
\eeq 
Using this method, we find
the terahertz pulse that detects any flipped orbital of the
$N$-bit data register. The advantage of this refinement is that
the computational resources needed increase only by a factor of
the number of constituent states in the wave packet register.

We now find the optimal terahertz pulse that will decode any
flipped state in a six state Rydberg data register.  The register consists of
$np$ states of cesium, with $n$ from $24-29$.  Population in the flipped orbital
is amplified by the diffusion of probability density from the adjacent states.  
This is an
example of the implementation of Grover's search algorithm, where information
is stored in states with differing phases, and a marked bit is amplified by
``quantum diffusion''~\cite{grover}.
The outer states $n=24$ and $n=29$ are therefore not included in the optimization.
The universal decoding
pulse and its effect on a wave packet with different marked bits is
shown in Fig.~\ref{fig3}.
After the pulse, the wave packet population
is distributed so that the flipped state is clearly amplified.  
\section{CONCLUSIONS}
In conclusion, we have designed a terahertz pulse to implement a search
algorithm on a quantum data register.  Phase information stored in a
Rydberg wave packet was optimally retrieved through the interaction with
the pulse.  Careful attention was paid to the smooth switch on and 
switch off of the THz pulse.  We also show that it is possible to design an 
optimal pulse that can achieve not only a desired target state of an atom, 
but also implement a desired algorithm.

To our knowledge, this is the first time that optimal control
theory has been applied to the terahertz control of a quantum system.  This
theoretical study motivates the experimental design and control of terahertz
frequency pulses.  Beyond quantum control, these results point to the
possibilities of using Rydberg atoms as quantum computers, and terahertz pulses
to implement quantum algorithms.

\section*{ACKNOWLEDGEMENTS}
It is a pleasure to acknowledge discussions and help from Professors D.~J.~Tannor,
A.~M.~Bloch, V.~F.~Krotov and V.~Malinovsky.  C.~R. acknowledges support
from the Fellows and Visitor's program of the Center for Ultrafast
Optical Science at the University of Michigan.  This research was 
supported by the National Science Foundation grant 9987916.

\end{document}